\documentclass[twocolumn,showpacs,preprintnumbers,amsmath,amssymb]{revtex4}
\usepackage{graphicx}% Include figure files
\usepackage{dcolumn}% Align table columns on decimal point
\usepackage{bm}% bold math

\begin{document}

%\preprint{APS/123-QED}

\title{The distance modulus determined from Carmeli's cosmology fits \\ the accelerating universe data of the high-redshift type Ia \\ supernovae without dark matter}% Force line breaks with \\

\author{John G. Hartnett}
 \email{john@physics.uwa.edu.au}
\affiliation{School of Physics, The University of Western Australia\\35 Stirling Hwy, Crawley 6009 WA Australia}%

% \homepage{}
%\affiliation{

\date{\today}% It is always \today, today,

\begin{abstract}
The velocity of the Hubble expansion has been added to General Relativity by Moshe Carmeli and this resulted in new equations of motion for the expanding universe.  For the first time the observational magnitude-redshift data derived from the high-$z$ supernova teams has been analysed in the framework of the Carmeli theory and the fit to that theory is achieved without the inclusion of any dark matter.  Best fits to the data yield an averaged matter density for the universe at the present epoch $\Omega_{m} \approx 0.021$, which falls well within the measured values of the baryonic matter density. And the best estimate of $\Omega_{\Lambda} + \Omega_{m} \approx 1.021$ at the present epoch. The analysis also clearly distinguishes that the Hubble expansion of the universe is speed-limited. 
\end{abstract}

\maketitle

\section{\label{sec:Intro}Introduction}

The metric \cite{Behar2000,Carmeli2002a} used by Carmeli in a generally covariant theory (Cosmological General Relativity) extends the number of dimensions of the universe by the addition of a new dimension -- the radial velocity of the galaxies in the Hubble flow. The Hubble law is assumed as a fundamental axiom for the universe and the galaxies are distributed accordingly. 

In determining the large scale structure of the universe the usual time dimension is neglected as observations are taken over such a short time period compared to the motion of the galaxies in the expansion. In this case, the usual time coordinate ($t$) does not need to be considered. This leaves only four dimensions to be considered -- three of space and one of velocity. In general, however, the new 5D cosmology developed by Carmeli contains all of general relativity as a subset. All of the results in general relativity that are experimentally supported are also obtained in Cosmological General Relativity (CGR) \cite{Carmeli2002b}. 

When discussing the motions of stars and gases in galaxies the five dimensional cosmology must be used. Both the usual geodesic equation and a new phase space equation have been derived using CGR. From this Carmeli was able to derive a Tully-Fisher like equation \cite{Carmeli1998}, which indicates that the existence of halo dark matter is not necessary to be assumed in spiral galaxies.

It is then worth seeing to what extent the Carmeli cosmological theory fits the observed data. This paper fits the theory to the observed magnitude-redshift data taken from the high-redshift type Ia supernovae observation teams. From this analysis on the largest scales of the universe no dark matter is necessary when the Carmeli model is assumed.   

\subsection{\label{sec:CGR}Cosmological General Relativity}

Here the CGR theory is considered using a Riemannian four-dimensional presentation of gravitation in which the coordinates are those of Hubble, i.e. distance and velocity, or more precisely, proper distances as measured by the Hubble law and the measured redshifts of galaxies. This results in a phase space equation where the observables are redshift and distance. The latter may be determined from the high-redshift type Ia supernovae (SNe Ia) observations. 

\subsection{\label{sec:lineelem}Line element}

In the case considered here the line element is
\begin{equation} \label{eqn:metric}
ds^{2}=\tau^{2}dv^{2}-e^{\xi}dr^{2}-R^{2}(d\theta^{2}+sin^{2}\theta d\phi^{2})
\end{equation}

where $ dr^{2}=(dx^1)^{2}+(dx^2)^{2}+(dx^3)^{2}$, $\xi$ and $R$ are functions of $v$ and $r$ alone and comoving co-ordinates $x^{\mu}=(x^{0},x^{1},x^{2},x^{3})=(\tau v ,r,\theta,\phi)$ are used. The new dimension ($v$) is the radial velocity of the galaxies in the expanding universe, and is not the time derivative of the distance. The parameter $\tau$, the Hubble-Carmeli time constant, is a constant for all observers at the same epoch and its reciprocal (designated $h$) is the Hubble `constant' measured in the limit of zero gravity and zero redshift, which is only approximately the Hubble constant $H_{0}$. 

Equation  (\ref {eqn:metric})  then represents a curved \textit{spacevelocity} which, like in general relativity, may be represented by a four-dimensional Riemannian manifold with a metric $g_{ \mu \nu}$ and a line element $ds^{2}=g_{ \mu \nu}dx^{\mu}dx^{\nu}$. This differs from general relativity in that here the $x^{0}$ coordinate is velocity-like instead of time-like as is the case of $x^{0} = ct$, where $c$ is the speed of light and a universal constant. The parameter $t$ is the time coordinate. In this theory $x^{0} = \tau v$, where $\tau$ is also a universal constant. The other three coordinates $x^{k}, k = 1,2,3$, are space-like, as in general relativity.

The line element represents a spherically symmetric isotropic but not necessarily homogeneous universe. The observations are made at a definite time and therefore $dt = 0$ does not appear in (\ref{eqn:metric}).  

\subsection{\label{sec:field}Field equations}

In CGR, as in general relativity, one equates geometry to physics. In this theory Einstein's field equations 
\begin{equation} \label{eqn:fieldeqn}
G_{\mu \nu}= R_{\mu \nu} - \frac {1} {2} g_{\mu \nu}R = \kappa T_{\mu \nu}
\end{equation}
are modified.

The energy-momentum tensor ($T_{\mu \nu}$) takes on a different physical meaning. The coupling constant ($\kappa$) that relates the geometry $G_{\mu \nu}$ in (\ref {eqn:fieldeqn}) to the energy terms $T_{\mu \nu}$ is also different. However the form of (\ref {eqn:fieldeqn}) is exactly the same as in general relativity, but with $\kappa = 8 \pi k/\tau^{4}$ and $ k = G \tau^{2}/c^{2}$ where $G$ is Newton's gravitational constant. Therefore $\kappa = 8 \pi G /c^{2} \tau^{2}$. 

The correspondence with general relativity is easily seen by the substitutions $c \rightarrow \tau$ and $t \rightarrow v$. In this new theory the energy-momentum tensor $T^{\mu \nu}$ is constructed with these substitutions. As usual  $T^{\mu \nu}= \rho u^{\mu} u^{\nu}$, where where $\rho$ is the average mass/energy density of the universe  and $u^{\mu} = dx^{\mu}/ds$ is the four-velocity. 

In general relativity $T_{0}^{0}=\rho$. In Newtonian gravity the potential function is defined by the Poisson equation $\nabla^{2}\phi=4 \pi G \rho$. Where $\rho = 0$ the vacuum Einstein field equations are usually solved in general relativity and similarly Laplace's equation in the Newtonian theory. 

These are valid solutions but in cosmology there never exists the situation where the density $\rho$ is zero because the universe always contains matter and energy. So in order to equate the rhs of (\ref {eqn:fieldeqn}) to zero Carmeli took $T_{0}^{0}\neq \rho$ but $T_{0}^{0} = \rho_{eff} = \rho - \rho_{c}$, in the appropriate units. Here $\rho_{c}$ is the critical or  ``closure'' density and in this model $\rho_{c} = 3/8 \pi G \tau^{2}\approx 10^{-29} \; g.cm^{-3}$. Therefore in CGR $T^{\mu \nu}= \rho_{eff} u^{\mu} u^{\nu}$. 

The result is that we can view the universe, in \textit{spacevelocity}, or phase space as being stress free when the matter density of the universe is equal to the critical density. That is, the effective density $\rho_{eff}= 0$. This then gives us the analogous situation to that in the Newtonian and Einsteinian theories. Besides the assumption of the universality of the Hubble Law, this is the second fundamental assumption in this cosmology.  

\subsection{\label{sec:phasespace}Phase space equation}

In \textit{spacevelocity} the null condition $ds = 0$ describes the expansion of the universe. In the limit of no matter or gravity (i.e. $e^{\xi}= 1 $) the null condition yields $dr/dv = \pm \tau$, which when integrated with appropriate initial conditions results in $r = \tau v$ in an expanding universe. This can be rewritten as $v = h r \approx H_{0} r$, the Hubble law in the zero-gravity limit.  

It follows from (\ref {eqn:metric}) for a spherically symmetric isotropic distribution of matter where spherical spatial coordinates ($r,\theta, \phi$) are used and taking into account $ d\theta = d\phi = 0 $ (the isotropy condition) that
\begin{equation} \label{eqn:4Dmetric}
\tau^{2}dv^{2}-e^{\xi}dr^{2}=0
\end{equation}
which results in
\begin{equation} \label{eqn:4Dmetricderiv}
\frac{dr} {dv} = \pm \tau e^{-\xi/2}.
\end{equation}
The positive sign is chosen for an expanding universe, which was solved in \cite{Carmeli2002a,Carmeli2002b}. 

\subsection{\label{sec:soln}Solution to field equations}

Carmeli found a solution to the resulting field equations \cite{Carmeli2002a}, with the necessary condition that $R'> 0$, as 
\begin{equation} \label{eqn:simple}
R = r 
\end{equation}
and 
\begin{equation} \label{eqn:soln1field2}
e^{\xi}= \frac{{1}}{1 + f(r)}
\end{equation}
where $f(r)$ is an arbirary function of $r$ and satisfies $f(r) + 1 > 0$. The solution is where 
\begin{equation} \label{eqn:fvalue1}
f(r) = \frac {1-\Omega}{c^{2}\tau^{2}}r^{2},
\end{equation}
and $\Omega = \rho/\rho_{c}$. The matter density $\Omega$ is assumed to be the smoothed average density for matter that is evenly distributed throughout the universe. However because we look back through past epochs the density is a function of the redshift $z$.  

By substituting (\ref{eqn:soln1field2}) with (\ref{eqn:fvalue1}) into (\ref {eqn:4Dmetricderiv}) we get Carmeli's result
\begin{equation} \label{eqn:phasespacederiv}
\frac{dr}{dv}= \tau \sqrt{ 1+(1-\Omega) \frac{r^2}{c^2 \tau^2}},
\end{equation}
where the positive solution has been chosen for an expanding universe.

Equation (\ref {eqn:phasespacederiv}) may be integrated exactly to get 
\begin{equation} \label{eqn:phasespacesoln}
r(v)= \frac{c \tau}{\sqrt{1-\Omega}}\sinh \left( \frac{v}{c} \sqrt{1-\Omega } \right)
\qquad \forall \Omega.
\end{equation}
Thus (\ref{eqn:phasespacesoln}) may be written in terms of normalized or natural units $ r/c \tau $ and for arbitrary $z = v/c$,
\begin{equation} \label{eqn:phasespacesolnnatural}
\frac {r} {c \tau}= \frac {\sinh (z \sqrt{1-\Omega})} {\sqrt{1-\Omega}}.
\end{equation}
Considering the expansion of the universe it is clear that (\ref{eqn:phasespacesolnnatural}) describes a tri-phase expansion. Initially the universe is very dense and $\Omega > 1$ so the hyperbolic sine function becomes a normal trignometric sine function describing a decelerating expansion. Then the density $\Omega$ reaches unity and the rhs of (\ref{eqn:phasespacesolnnatural}) becomes equal to $z$ which describes a coasting stage. Finally the density decreases and $\Omega < 1$ as the universe continues to expand. Then the hyperbolic sine function represents an exponentially accelerating universe.

There are four symbols used in this paper for density expressed as a fraction of the critical density. The symbol $\Omega$ represents the matter density at any epoch defined by redshift $z$ and because we take $dt = 0$, $\Omega$ is therefore only a function of $z$. The symbol $\Omega_{m}$ represents the matter density at the present epoch and therefore is a constant on the time scale of any measurements used here. The symbol $\Omega_{b}$ specifically represents the baryonic matter density at the present epoch, which in this paper we show is identical with  $\Omega_{m}$. The symbol $\Omega_{\Lambda}$ is the vacuum or `dark' energy contribution to gravity and is a function of redshift $z$. In fact, because the cosmological constant ($\Lambda$) does not appear explicitly in Carmeli's cosmology this latter component is really a property of the metric. 

The structure of this paper is as follows: Section \ref{sec:density} discusses the variation of matter density with redshift and presents an equation, modified by the author \cite{Hartnett2004b} from the Carmeli equation, that describes proper distance as a function of redshift in the Carmeli cosmology. Section \ref{sec:Comparison} is where the new work begins. In this section, for the first time,  a magnitude-redshift relation is fitted to the data taken from the high-$z$ supernovae teams \cite{Knop2003, Riess2004, Tonry2003}, for a model that assumes flat \textit{spacevelocity}. In Section \ref{sec:density2} a magnitude-redshift relation is fitted to the same data but this time for a model that assumes curved \textit{spacevelocity}. Section \ref{sec:Hubble} calculates the value of $h = 1/\tau$. Section \ref{sec:darkenergy} calculates the value of $\Omega_{\Lambda}$ and the total $\Omega_{\Lambda} + \Omega$ as a function of redshift. The appendix calculates, in a more rigorous way than in Section \ref{sec:density2}, the effect of curved \textit{spacevelocity} on matter density as a function of redshift.
\section{\label{sec:density}Matter density verses redshift}

Now let us consider the density of matter as a function of redshift, $z$. Carmeli assumed  that the value of $\Omega$ in (\ref{eqn:phasespacesolnnatural}) is fixed and plotted curves as functions of redshift for various values of $\Omega$. See figure A4, on page 134 of \cite{Carmeli2002a}. 

In 1996 he predicted that the universe must be accelerating \cite{Carmeli1996} and simulated the form of the high redshift data of Riess \textit{et al} \cite{Riess1998}, published in 1998, which announced an accelerating universe following the observations of Garnavich \textit{et al} \cite{Garnavich1997} and Perlmutter \textit{et al} \cite{Perlmutter1997}. 

Therefore in 1998, Carmeli assumed a value of total matter (normal bayonic matter + dark matter) density  $\Omega = 0.245$,  which was the accepted value then for $\Omega_{m}$ the matter density at the present epoch. However, more correctly $\Omega$ varies as a function of redshift, $z$. 

Carmeli never fitted his theory to any of the high redshift SNe data with a least squares method or any other, nor did he determine a value for the matter density $\Omega_{m}$ himself.

\subsection{\label{sec:density1}Baryonic matter density}

The density of normal baryonic matter in the universe has generally come from considerations of big bang nucleosynthesis (BBN)\cite{Krauss1998, Ostriker1995}. The predicted primordial abundance of the light elements are assumed to have been produced in the big bang and not in stars. Hence a comparison between predicted abundances and abundances inferred from observed baryon-to-photon ratios are used to put an upper limit on the baryon density. 

Deuterium is believed to be only destroyed by stellar processing, therefore any observation of its interstellar or solar system abundance would put a lower bound on its primordial abundance since it is believed that it was all created in the BBN process. As a result a lower bound on deuterium translates into an upper bound on the density of baryons. Following this line of logic the observed interstellar value of $D/H \geq (1.6 \pm 0.2)\times 10^{-5}$ puts a limit on the baryon density $\Omega_{b}h^{2}\leq 0.027 {}(2 \sigma)$ \cite{Krauss1998}. An earlier determination produced $\Omega_{b}h^{2}\leq 0.015 \pm 0.005$ \cite{Ostriker1995}. (Here $\Omega_{b}$ is the baryonic density expressed as a fraction of the critical density and $h$ is the Hubble constant as a fraction of $100 \; km.s^{-1} Mpc^{-1}$ and not to be confused with $h = 1/\tau$ used in this paper.) Assuming a value of the Hubble parameter $h = 0.7$ these two sources put $\Omega_{b} \leq 0.055$ and $\leq 0.031$ respectively.

The aforementioned values of $\Omega_{b}$ agree well with the locally measured baryonic budget. One study yields a range $0.007 \leq \Omega_{b} \leq 0.041$ at $z\approx 0$ with a best guess of $\Omega_{b} \approx 0.021$ where a Hubble constant of $70 \; km.s^{-1} Mpc^{-1}$ was assumed \cite{Fukugita1998}.

Now, determinations of the total matter density (including dark matter), at high redshifts, have been made, for example, from X-ray measurements of the gas content in galaxy clusters combined with the total virial masses \cite{White1993}. The calculation of the virial masses in turn assumes that the equations describing the motions of the constituents in the clusters are correct and therefore gravitational potentials are assumed that generate masses larger than the baryonic matter content. Hence a significant dark matter content results. 

However it is well-known that mass-to-light ratios vary over all mass scales from galaxies to superclusters \cite{Wright1990}. This may be the result of an incorrect understanding of the physics which has resulted in incorrect dynamical masses. That conclusion is suggested by Carmeli's derivation of a Tully-Fisher type relation for galaxies \cite{Carmeli1998} and by a new post-Newtonian equation describing the dynamics of the stars and gases in spiral galaxies \cite{Hartnett2004a}. The latter produces the galaxy rotation curves, without any dark matter, where previously it was necessary to assume that they are evidence for halo dark matter. 

Therefore the application of the Carmeli theory to the very important problem of galaxy cluster dynamics is necessary to see if it also eliminates dark matter on that scale. Eventually it is necessary that the Carmeli theory also determine cluster masses from the virial theorem, observed X-ray temperatures, and from weak gravitational lensing with results consistent with each method and with the observed baryonic matter density. However such an exercise is beyond the scope of this paper. 

The question that remains for this paper to answer is ``Do the high-$z$ SNe Ia data fit this model without the need to assume the existence of dark matter?'' In future research, other issues can be addressed, for example, to explain the much higher than expected dynamical masses of clusters at high redshift, and the shape of the blackbody spectrum of the cosmic microwave background as well as the acoustic peaks in its spatial spectrum. 

\subsection{\label{sec:density2}Matter density at the present epoch}

This paper makes no determinations about the matter density, dark, baryonic or otherwise at high redshift except the following, with a justification in section \ref{sec:darkenergy}.
\begin{equation} \label{eqn:densityeqn}
\Omega = \Omega_{m}(1+z)^{3},
\end{equation}                           
where $\Omega_{m}$ is then the averaged matter density, at the present epoch, expressed as a fraction of the critical or ``closure'' density. $\Omega_{m}$  can be considered to be constant on the time scale of any measurements used here. Equation (\ref {eqn:densityeqn}) results from the fact that as the redshift increases the volume changes as $(1 + z)^{-3}$ assuming flat Euclidean space.  

In the standard Friedmann-Lemaitre (F-L) theory the equivalent expression is
\begin{equation} \label{eqn:stdmodeldensityeqn}
\Omega = \Omega_{m}a^{-3}/H(a)^{2},
\end{equation}
where $a = (1+z)^{-1}$ the scale factor (because at the present epoch, as usual, $a$ is assumed to be unity) and $H(a)$ is the Hubble term which quantifies the curvature of the expansion and defined $H^{2} \equiv (\dot{a}/a)^{2}$, where the dot is the time derivative.

In this theory the curvature results from the inclusion of the new velocity dimension and the implicit assumption of the universality of the Hubble law. This introduces curvature through the concept of \textit{spacevelocity}, which is discussed later. (See section \ref{sec:density2} and the appendix. The time derivative of a scale factor is not relevant.) 

In the first instance I assume a flat  \textit{spacevelocity} model in (\ref{eqn:densityeqn}), which is identical to (\ref{eqn:stdmodeldensityeqn}) where $H(a) = 1$. This would correspond to a dust dominated spatially flat universe in the standard theory.

Now the value of the baryonic matter density commonly cited is $\rho = 3 \times 10^{-31} \; g.cm^{-3}$. If this is assumed to be the total matter density then $\Omega_{m} \approx 0.03$ in (\ref {eqn:densityeqn}). 

Substituting (\ref {eqn:densityeqn}) into (\ref {eqn:phasespacesolnnatural}) we get 
\begin{equation} \label{eqn:phasespacesolnnaturald}
\frac {r} {c \tau}= \frac {\sinh \left(z \sqrt{1-\Omega_{m}(1+z)^3 }\right)} {\sqrt{1-\Omega_{m}(1+z)^3}}.
\end{equation}

It has been shown \cite{Hartnett2004b} that (\ref {eqn:phasespacesolnnatural}) with $\Omega = 0.245$ (which Carmeli initially assumed) and (\ref {eqn:phasespacesolnnaturald}) with $\Omega_{m} = 0.03$ are practically identical over the redshift range $ 0 < z < 1$. The difference between the two equations over the domain of the measurements is much less significant than the fit to the data. 

Once it is shown that this theory fits the actual observational data it follows that this effectively eliminates the need for the existence of dark matter on the cosmic scale.

\section{\label{sec:Comparison}Comparison with high-z type Ia supernovae data}

In order to compare (\ref{eqn:phasespacesolnnaturald}) with the data from the high redshift SNe Ia teams the proper distance is converted to magnitude as follows.
\begin{equation} \label{eqn:lumindistance}
m(z) = \mathcal{M} + 5log \mathcal{D}_{L}(z;\Omega), 
\end{equation}
where $\mathcal{D}_{L}$ is the ``Hubble constant free'' luminosity distance. $\mathcal{D}_{L}$ is related by
\begin{equation} \label{eqn:magnitude}
\mathcal{D}_{L}(z;\Omega) = (1+z) \frac {r} {c \tau} 
\end{equation}
using (\ref{eqn:phasespacesolnnaturald}) and 
\begin{equation} \label{eqn:scriptem}
\mathcal{M} = 5log(c \tau) + 25 + M_{B} + a.
\end{equation}

Here $\mathcal{M}$ incorporates the various parameters that are independent of the redshift, $z$.  The parameter $M_{B}$ is the absolute magnitude of the supernova at the peak of its light-curve and the parameter $a$ allows for any uncompensated extinction or offset in the mean of absolute magnitudes or an arbitrary zero point. The absolute magnitude then acts as a ``standard candle'' from which the luminosity and hence distance can be estimated. 

The value of $M_{B}$ need not be known, neither any other component in $\mathcal{M}$, as $\mathcal{M}$ has the effect of merely shifting the fit curve (\ref{eqn:magnitude}) along the vertical magnitude axis. 

By choosing the value of $\tau = 4.2 \times 10^{17} \, s = 13.3 \; Gyr$, which is the reciprocal of the chosen value of the Hubble constant in the gravity free limit $h = 73.54 \; km.s^{-1} Mpc^{-1}$ (see section \ref{sec:Hubble}) $\mathcal{M} = 43.06 + M_{B} + a$. In practice, where the distance modulus ($m-M_{B}$) is used in the curve fits, $a$ is an arbitrary free parameter.

It has been one of the goals of cosmology to determine if the Hubble expansion is speed limited. That is, to answer the question of whether the cosmological expansion is governed by the relativistic Doppler effect or not. If it isn't then in (\ref{eqn:phasespacesolnnaturald}) $v/c = z$ (as written), but if it is then as $v$ approaches $c$ it is necessary to replace $z$ in (\ref{eqn:phasespacesolnnaturald}) with $v/c = ((1+z)^{2}-1)/((1+z)^{2}+1)$. So by fitting to the data of the high-$z$ SNe Ia it is possible to test this experimentally. Behar and Carmeli \cite{Behar2000} have asserted that latter (speed-limited version) is the correct form, so this then becomes an experimental test of the prediction.

\begin{figure}
\includegraphics[width = 3.5 in]{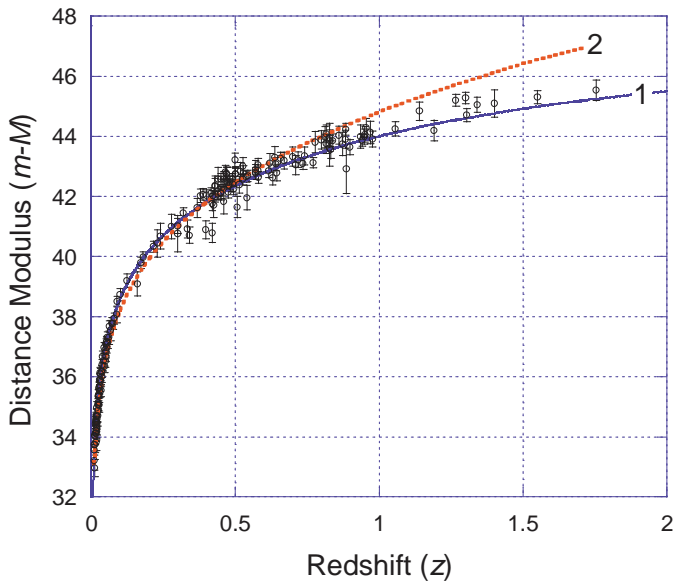}
\caption{\label{fig:fig1} (Color online) Data taken from Table 5 of Riess \textit{et al}. Curve 1 (purple solid line) is the best fit and respresents the speed limited model. Curve 2 (orange dots) is the non speed limited model. Both curves use the same value of $\Omega_{m} = 0.02$. In order to make curve 2 pass through the highest data points the required mass density $\Omega_{m}$ must be larger than $0.10$. However such values are unphysical}
\end{figure}

In figure \ref{fig:fig1} (\ref{eqn:lumindistance}) has been plotted against the distance modulus ($m-M_{B}$) data taken from Table 5 of Riess \textit{et al} \cite{Riess2004}. Curve 1 is the best fit to the data using (\ref{eqn:phasespacesolnnaturald}) with $v/c = ((1+z)^{2}-1)/((1+z)^{2}+1)$, the speed-limited model and curve 2 is (\ref{eqn:phasespacesolnnaturald}) with $v/c = z$ the non-speed-limited model. This immediately answers the question. The Hubble expansion, as seen from our position as observer, is speed-limited, by the usual relativistic Doppler effect, which confirms the theoretical prediction.

It can be seen that curve 2 departs from curve 1 slightly above redshift, $z = 0.5$. Both curves have been plotted with a value of $\Omega_{m} = 0.02$, which is the best fit for curve 1. However different `best fit' offsets ($a$) have resulted. For curve 1, $a = 0.41$ and for curve 2, $a = 0.011$. There are a total of $N = 185$ data and the residual for the speed-limited version of (\ref{eqn:phasespacesolnnaturald}) is $R = 0.9960$, with $\chi^{2} = 19.94$ and hence $\chi^{2}/N = 0.1078$. The residual for the non-speed-limited version of (\ref{eqn:phasespacesolnnaturald}) is $R = 0.9927$, with $\chi^{2} = 37.465$ and hence $\chi^{2}/N = 0.2036$.

In order to make curve 2 pass through the highest data points the requirement on the mass density is $\Omega_{m} \geq 0.10$, which is unphysical assuming only baryonic matter. As mentioned in section \ref{sec:density}, the local baryonic matter budget has been measured in the range $0.007 \leq \Omega_{b} \leq 0.041$ \cite{Fukugita1998} with a best guess of $\Omega_{b} \approx 0.021$. The latter is consistent with the best fit from this data set.  Therefore, in the following analysis, $\Omega_{m} = \Omega_{b}$ i.e. no dark matter needs to be included. Also only the speed-limited form of (\ref{eqn:phasespacesolnnaturald}) will be used. That is
\begin{equation} \label{eqn:speedlimited}
\frac {r} {c \tau}= \frac {\sinh \left(\varsigma \sqrt{1-\Omega_{m}(1+z)^3 }\right)} {\sqrt{1-\Omega_{m}(1+z)^3}}   
\end{equation}
where $ \varsigma = ((1+z)^{2}-1)/((1+z)^{2}+1)$.
\begin{figure}
\includegraphics[width = 3.5 in]{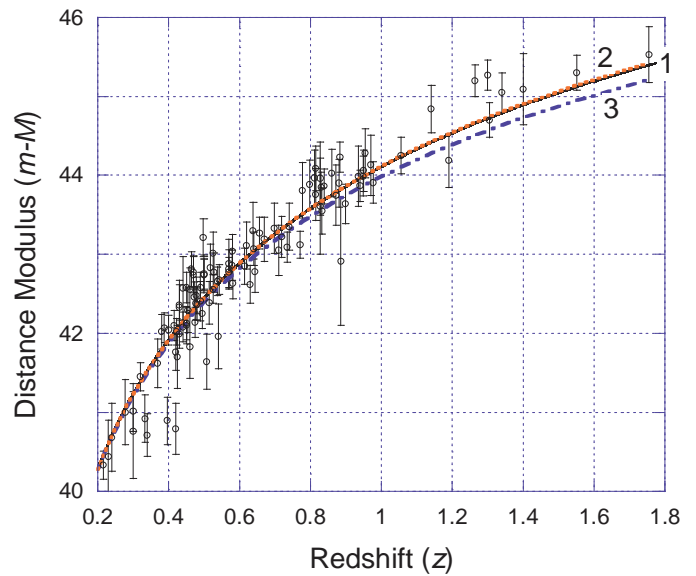}
\caption{\label{fig:fig2} (Color online) Data taken from Riess \textit{et al} for $z > 0.2$. Curve 1 (black solid line) is the best fit and respresents (\ref{eqn:lumindistance}) with the speed-limited (\ref{eqn:speedlimited}). Curve 2 (orange dots) represents (\ref{eqn:lumindistance}) with $\Omega = \Omega_{m} = 0$ and curve 3 (purple dot dash) represents (\ref{eqn:lumindistance}) with $\Omega = 1$}
\end{figure}

Figure \ref{fig:fig2} shows the data of figure \ref{fig:fig1} but for $ z > 0.2$. Curve 1 (solid line) is the best fit curve over the range of the selected data and also consistent with measured baryonic density i.e. $\Omega_{m} \geq 0.007$. Instead of $\Omega_{m} = 0.02$, the best fit curve requires $\Omega_{m} = 0.007 \pm 0.050$ (statistical). However, the value of $\Omega_{m}$ may range up to $0.04$ with little change in residuals. A slightly increased value of $a = 0.53$ results compared with the fit in figure \ref{fig:fig1}. This may indicate a small $0.12$ additional extinction over the low redshift data. Here $N = 106$, the residual $R = 0.9946$, $\chi^{2} = 17.107$ and hence $\chi^{2}/N = 0.1614$. Curve 2 (dots) represents (\ref{eqn:lumindistance}) with $\Omega = \Omega_{m} = 0$ and curve 3 (dot dash) represents (\ref{eqn:lumindistance}) with $\Omega = 1$. Curve 3 represents a universe where the matter density is always critical regardless of epoch.

\begin{figure}
\includegraphics[width = 3.5 in]{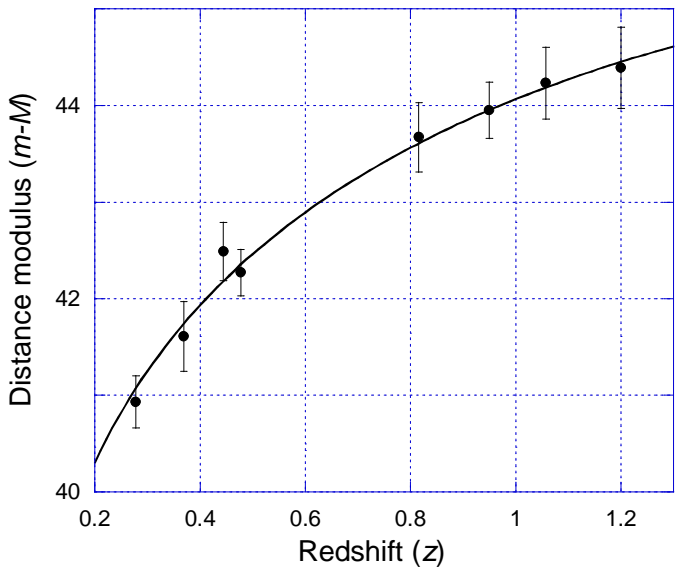}
\caption{\label{fig:fig3} Data taken from Tonry \textit{et al}. The curve (solid line) is the best fit and respresents (\ref{eqn:lumindistance}) with the speed-limited (\ref{eqn:speedlimited})}
\end{figure}

Figure \ref{fig:fig3} shows the 8 new SNe Ia data from Tonry \textit{et al} \cite{Tonry2003} for sources with $z > 0.2$. The resulting best fit is shown, with $\Omega_{m} = 0.073 \pm 0.098$ (statistical), $a = 0.56$, the residual $R = 0.9938$, $\chi^{2} = 0.1460$ and hence $\chi^{2}/N = 0.0183$. Tonry \textit{et al} used 4 methods of analysis and averaged the results. The data shown in figure \ref{fig:fig3} is drawn from their Table 6, column 4. 

\begin{figure}
\includegraphics[width = 3.5 in]{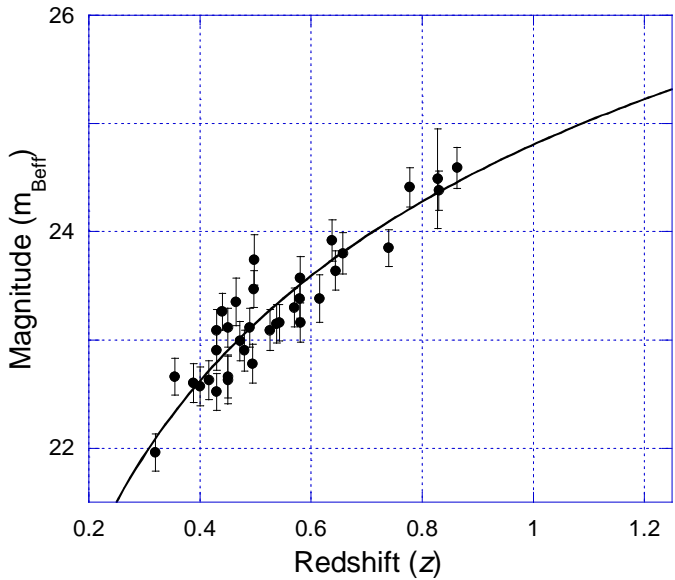}
\caption{\label{fig:fig4} Data taken from Knop \textit{et al} for $z > 0.2$. The curve (solid line) is the best fit and respresents (\ref{eqn:lumindistance}) with the speed-limited (\ref{eqn:speedlimited})}
\end{figure}

Figure \ref{fig:fig4} shows 30 type SNe Ia data (which includes 11 new SNe) from Knop \textit{et al} \cite{Knop2003} for sources with $z > 0.2$. In this case the magnitude has been plotted against redshift. These data are taken from column 4 of Tables 3 and 4 in Knop \textit{et al}.  As with the other data various corrections have been applied, including $K$-correction, Galactic extinction, and light-curve stretch correction. (See references for details). In figure \ref{fig:fig4}, the host galaxy corrected data (column 5 from (Knop \textit{et al})) was not used as it added a large scatter without significant improvement to the fit. The resulting best fit is shown, with $\Omega_{m} = 0.019 \pm 0.137$ (statistical), the residual $R = 0.9217$, $\chi^{2} = 1.9323$ and hence $\chi^{2}/N = 0.0644$. Because the effective apparent magnitude $m_{Beff}$ was used instead of distance modulus the offset is $M_{B}+ a = -18.760$. 

By taking the statistically weighted average of the matter density $\Omega_{m}$ obtained from each curve fit in figures \ref{fig:fig2} - \ref{fig:fig4}, we get an average value 
\begin{equation} \label{eqn:statav}
\Omega_{m} = \sum_{i} \frac{ \Omega_{mi}}{\sigma_{i}^{2}}/\sum_{i}\frac{1}{\sigma_{i}^{2}} = 0.021 \pm 0.042,
\end{equation}
where $\Omega_{mi}$ are the best fit values from the three data sets used and $\sigma_{i}$ is the standard error for $\Omega_{mi}$.

\section{\label{sec:density2}Curved \textit{spacevelocity}}

In section \ref{sec:density} on matter density it was assumed that the density could be described by the function in (\ref{eqn:densityeqn}) that assumes the universe is essentially spatially Euclidean. This assumption is justified in section \ref{sec:darkenergy}. However it is obvious that besides for small $z$ and where $\Omega = 1$, \textit{spacevelocity} is not flat in general and is expected to modify the expression in (\ref{eqn:densityeqn}). When $\Omega > 1$ it is curved and closed and when $\Omega < 1$ it is curved and open. Equation (\ref{eqn:densityeqn}) relates the matter density at any epoch to the present epoch value ($\Omega_{m}$) in Euclidean space or flat \textit{spacevelocity}. 

Rewriting (\ref{eqn:4Dmetricderiv}) in an expanding universe we get
\begin{equation} \label{eqn:4Dmetricderivrz}
\frac {1} {c \tau} \frac{dr} {dz} = e^{-\xi/2}.
\end{equation}
The result is the gradient of the spatial co-ordinate $r$ with respect to the redshift $z$, normalized by the Hubble length $c \tau$. 

Therefore it is clear from (\ref{eqn:4Dmetricderivrz}) that $e^{\xi/2} = 1$ under the conditions that produce the Hubble law in the limit of weak gravity or for flat \textit{spacevelocity}. It can be seen from (\ref{eqn:soln1field2}) and (\ref{eqn:fvalue1}) that $e^{\xi/2} = 1$ when $\Omega = 1$. Therefore we can use $e^{\xi/2}$ to define the curvature of \textit{spacevelocity} as a function of redshift $z$. Because the density scales as the inverse cube of the radial coordinate and since the radial coordinate scales differentially as  $e^{-\xi/2}$, then $e^{3\xi/2}$ may be a good estimate of how the density scales with redshift $z$. 

See the appendix for a more rigorous approach. The more rigorous approach yields an integral that can only be approximated and results in a transcendental equation which cannot be used in the curve fits. Whereas the former approach results in an analytic function that can be used in the curve fits. 

By combining (\ref{eqn:densityeqn}) with (\ref{eqn:soln1field2}) and (\ref{eqn:fvalue1}), for $e^{\xi/2} \neq 1$ we can write
\begin{equation} \label{eqn:densitycurved}
\Omega = \frac {\Omega_{m} (1+z)^{3}} {\left (1+(1-\Omega)\left(\frac {r} {c \tau}\right)^{2}\right)^{3/2}}.
\end{equation}
When $\Omega = 1$ in the denominator of (\ref{eqn:densitycurved}) we recover (\ref{eqn:densityeqn}). To get the density $\Omega$ as a function of $z$ we substitute $r/c \tau$ from (\ref{eqn:speedlimited});
\begin{equation} \label{eqn:densitycurved2}
\Omega = \frac {\Omega_{m} (1+z)^{3}} {\left(1 + sinh^{2}\left(\varsigma \sqrt{1 - \Omega_{m} (1+z)^{3}} \right)\right)^{3/2}},  
\end{equation}
where $\varsigma = ((1+z)^{2}-1)/((1+z)^{2}+1)$.

\begin{figure}
\includegraphics[width = 3.5 in]{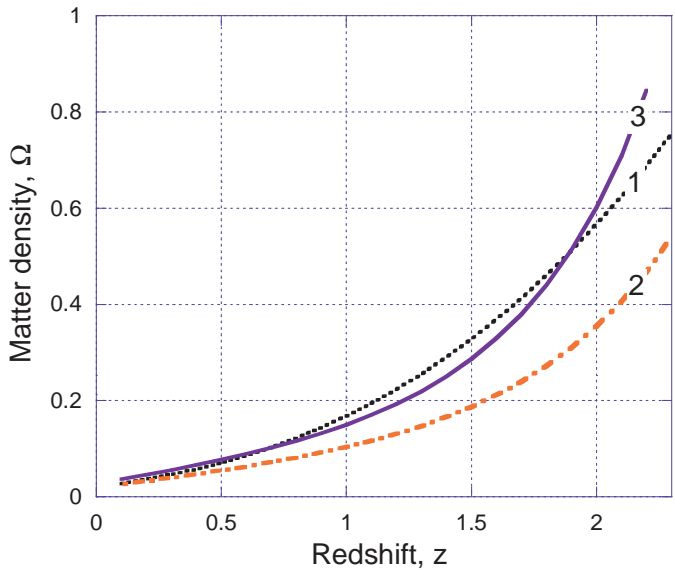}
\caption{\label{fig:fig5} (Color online) Matter density as a function of redshift $z$ for the two models. Curve 1 (black dots) is the flat \textit{spacevelocity} density from (\ref{eqn:densityeqn}) with $\Omega_{m} = 0.021$ taken from (\ref{eqn:statav}). The curved \textit{spacevelocity} density from (\ref{eqn:densitycurved2}) with $\Omega_{m} = 0.021$ is represented by curve 2 (orange dot dash), and with $\Omega_{m} = 0.027$ by curve 3 (purple solid line)}
\end{figure}

This results in a new density function which is compared with (\ref{eqn:densityeqn}) in figure \ref{fig:fig5}. Curve 1 is the flat \textit{spacevelocity} density from (\ref{eqn:densityeqn}) with $\Omega_{m} = 0.021$ taken from (\ref{eqn:statav}). Curve 2 is for the curved \textit{spacevelocity} density from (\ref{eqn:densitycurved2}) with the same value of $\Omega_{m} = 0.021$. Note the curves 1 and 2 are equivalent for $z < 0.4$.  For small $z$ we recover (\ref{eqn:densityeqn}) from (\ref{eqn:densitycurved2}). Because of the condition $f(r) + 1 > 0$ equation (\ref{eqn:densitycurved2}) is valid for all $z < 4$ provided $\Omega_{m} < 0.03$. 

The density from (\ref{eqn:densitycurved2}) is then substituted into (\ref{eqn:speedlimited}) replacing $\Omega_{m}(1+z)^{3}$ to get a new expresssion for $r/c \tau$ with only one free parameter $\Omega_{m}$. This was then used in the same curve fits in section \ref{sec:Comparison} (figures \ref{fig:fig1} - \ref{fig:fig4}) with the following results. 

In each case the curve fits were improved by the new density model except for the figure \ref{fig:fig2} fit, which had a marginally worse $\chi^{2}$ statistic. For the curve 1 of figure \ref{fig:fig1} the $\Omega_{m} = 0.02$ fit had a reduced but the same $R = 0.9960$. This resulted in a slight change in the offset $a = 0.41$. For figure \ref{fig:fig2} the best fit yielded $\Omega_{m} = 0.016 \pm 0.115$ (statistical) with offset $a = 0.56$ and residual $R = 0.9945$, $\chi^{2} = 17.208$ and hence $\chi^{2}/N = 0.1623$. For figure \ref{fig:fig3} the best fit yielded $\Omega_{m} = 0.020 \pm 0.097$ (statistical) with offset $a = 0.56$ and residual $R = 0.9928$, $\chi^{2} = 0.1684$ and hence $\chi^{2}/N = 0.0211$. For figure \ref{fig:fig4} the best fit yielded $\Omega_{m} = 0.029 \pm 0.175$ (statistical) with offset $M_{B}+ a = -18.76$ and residual $R = 0.9218$, $\chi^{2} = 1.9317$ and hence $\chi^{2}/N = 0.0644$. 

Finally by taking the weighted average as before
\begin{equation} \label{eqn:statav2}
\Omega_{m} = \sum_{i} \frac{ \Omega_{mi}}{\sigma_{i}^{2}}/\sum_{i}\frac{1}{\sigma_{i}^{2}} = 0.020 \pm 0.068.
\end{equation}

The new density function (\ref{eqn:densitycurved2}) yields the same result as the previous model. However in the Appendix it is shown that a more rigorous approach indicates that the curved \textit{spacevelocity} model matter density approximates the density of the flat \textit{spacevelocity} model with only a 1.28 multiplying factor for $z<2$. This means the best fit derived value $\Omega_{m} = 1.28 \times 0.021 = 0.027$. However the difference in the fits using $\Omega_{m} = 0.021$ or $0.027$ is extremely small and only apparent for the few data in the region $z >1$, but much less than their associated error bars.

\section{\label{sec:Hubble}Hubble parameter}

By inverting (\ref{eqn:speedlimited}), multiplying both sides of the resulting equation by $v/c$ and using the expression $v = H_{0}r$ we get the following
\begin{equation} \label{eqn:hubble}
H_{0} = h \frac {\varsigma \sqrt{1-\Omega}} {\sinh \left(\varsigma \sqrt{1-\Omega }\right)}.
\end{equation}         
with the density $\Omega$ taken from either (\ref{eqn:densityeqn}) or (\ref{eqn:densitycurved2}). 

As a result we have an expression which indicates that the Hubble parameter $H_{0}$ is scale length related and a function of the universal Hubble constant ($h$). It has been common to find figures cited in the literature that indicate `long' and `short' scales for $H_{0}$. Studies with the Hubble Space Telescope using classical Cepheid variables yielded $H_{0} = 80 \pm 17 \; km.s^{-1} Mpc^{-1}$ \cite{Freedman1994} whereas studies using SN Ia yielded $H_{0} = 67 \pm 7 \; km.s^{-1} Mpc^{-1}$ \cite{Riess1995}. See also \cite{Freedman2000}. 

It then follows from (\ref{eqn:hubble}) that $h \approx 73.54 \; km.s^{-1} Mpc^{-1}$ for $\Omega_{m} = 0.021$ and the flat \textit{spacevelocity} model (\ref{eqn:densityeqn}) when a value of $H_{0} = 70.00 \; km.s^{-1} Mpc^{-1}$ at $z = 1$ is assumed. In any case, the specific value of $h$ or it reciprocal $\tau$ cannot be determined fom the data fit. An independent method must be used to determine the precise value of $\tau$.

\section{\label{sec:darkenergy}Dark energy}

The vacuum or so-called `dark' energy parameter $\Omega_{\Lambda}$ does not appear explicitly in Carmeli's CGR.  Hence the term `dark' energy is a misnomer, and probably `vacuum' energy is more correct. In any case, it is really a property of the metric. Only by a comparison with the standard F-L models can an assignment be made \cite{Behar2000, Carmeli2002a}. 

\begin{figure}
\includegraphics[width = 3.5 in]{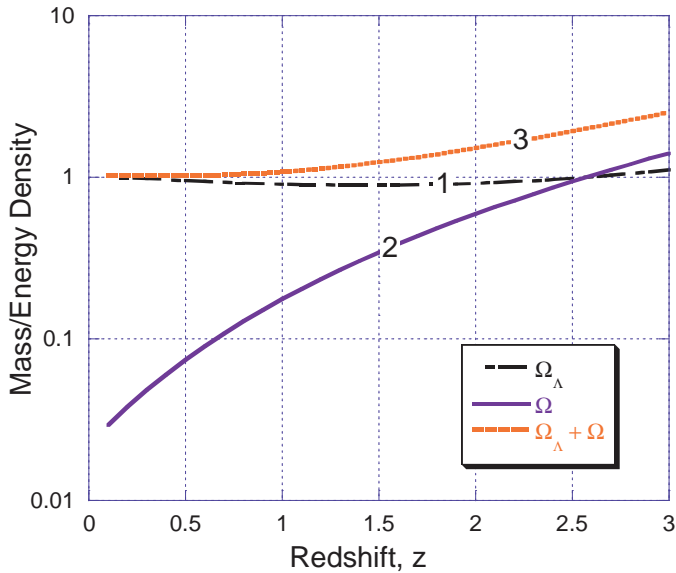}
\caption{\label{fig:fig6} (Color online) $\Omega_{\Lambda}$ (curve 1), $\Omega$ (curve 2) and the total mass/energy density $\Omega_{\Lambda} + \Omega$ (curve 3) as a function of redshift $z$ for the flat \textit{spacevelocity} density model with $\Omega_{m} = 0.021$}
\end{figure}

The vacuum energy density $\rho_{\Lambda} = \Lambda/8 \pi G$ (in CGR) $= 3H_{0}^{2}/8 \pi G$ (from the standard theory). Also in CGR the critical density $\rho_{c} = 3h^{2}/8 \pi G$.  Therefore $\Omega_{\Lambda} = \rho_{\Lambda}/\rho_{c} = (H_{0}/h)^2$ and it follows from (\ref{eqn:hubble}) that

\begin{equation} \label{eqn:Darkenergy}
\Omega_{\Lambda} =   \left(\frac{\varsigma \sqrt{1-\Omega}} {\sinh \left(\varsigma \sqrt{1-\Omega }\right)}\right)^2,
\end{equation} 
where the density $\Omega$ is taken from (\ref{eqn:densityeqn}) and the resulting $\Omega_{\Lambda}$, as a function of $z$, is shown in figure \ref{fig:fig6} for the flat \textit{spacevelocity} density model. Curve 1 and 2 are respectively the values of $\Omega_{\Lambda}$  and $\Omega$ with $\Omega_{m} = 0.021$. 

Curve 3 in fig. \ref{fig:fig6} shows the values for the total energy density  $\Omega + \Omega_{\Lambda}$  as a function of redshift, $z$, for the flat \textit{spacevelocity} model with $\Omega_{m} = 0.021$. The total density $\Omega + \Omega_{\Lambda} \approx 1$ for $z < 1$ and still remains close to unity up to $z = 2$. This means out to $z \approx 2$ the universe is quasi-Euclidean. 

Table \ref{tab:table1} shows the two models compared with $\Omega_{m} = 0.02, 0.03, 0.04$ which are all within the bounds of the measured baryonic matter density at the present epoch ($z \approx 0$). Fits from the two models (flat with $\Omega_{m} \approx 0.02$; curved with $\Omega_{m} \approx 0.03$) both result in approximately the same total density $\Omega + \Omega_{\Lambda} = 1.067$ at $z = 1$. 

For small $z$ the total density becomes
\begin{equation} \label{eqn:Darkenergysmz}
\Omega + \Omega_{\Lambda} \approx   (1+\Omega_{m})+3z\Omega_{m}.
\end{equation} 

It follows from (\ref {eqn:Darkenergysmz}) that for  $\Omega_{m} \approx 0.021$ at $z \approx 0$ the total density  $\Omega + \Omega_{\Lambda}\approx 1.021$. 

Furthermore now let us consider the time development of these densities in a qualitative sense only. From (\ref {eqn:Darkenergy}) it follows that as the universe expands the total density tends to the vacuum energy density  $\Omega_{\Lambda}\rightarrow 1$ because $\Omega_{m}\rightarrow 0$. This means a 3D spatially flat universe in a totally relaxed state. 

\begin{table}
\caption{\label{tab:table1}Mass and energy fractions at a redshift of $z = 1$ for flat and curved models and various values of $\Omega_{m}$}
\begin{ruledtabular}
\begin{tabular}{ccccccc}
 Density &flat &curved &flat& curved &flat &curved\\
\hline
$\Omega_{m}$& $0.02$ & 0.02 &0.03 &0.03 & 0.04 & 0.04  \\
$\Omega_{\Lambda}$ & 0.905 & 0.899 & 0.914 & 0.905 & 0.922 & 0.912 \\
$\Omega$ & 0.16 & 0.104 & 0.24 & 0.162 & 0.32 & 0.225 \\
$\Omega + \Omega_{\Lambda}$ & 1.065 & 1.003 & 1.154 & 1.067 & 1.242 & 1.137 \\
\end{tabular}
\end{ruledtabular}
\end{table}

\section{\label{sec:conclusion}Conclusion}

The 5D brane world of Moshe Carmeli has been has been applied to the accelerating expanding universe and the magnitude-redshift distance relation has been applied to the distance modulus data from the type Ia SNe measurements.  It has been found that considering only the evolution of baryonic matter density as a function of redshift, the resulting distance-redshift relation fits the data of the high-$z$ supernova teams without the need for any dark matter. 

Astronomers and cosmologists have wondered whether the expansion of the universe is speed-limited in the usual Doppler sense. Here it has also been shown that the expansion is, in fact, speed-limited.  This means that even though the galaxies are assumed fixed within the expanding space, energy cannot be transported faster than the speed of light in vacuum, i.e. photons arriving at an Earth detector are relativistically Doppler redshifted.

Considering the range set on the present epoch baryonic matter density of $0.007 \leq \Omega_{b} \leq 0.041$ \cite{Fukugita1998} with a best guess of $\Omega_{b} \approx 0.021$ and the fact that necessarily $\Omega_{m} > 0$, the best estimates of the matter density (which is only baryonic) from this analysis are $\Omega_{m} = 0.021_{-0.014}^{+0.020}$ with the flat \textit{spacevelocity} model. A similar result was obtained fitting a curved \textit{spacevelocity} model. 

Even though the statistically derived standard errors are large, it is only necessary that $\Omega_{m}$ be within the range of the locally measured baryonic matter density for the theory to fit the data. Therefore dark matter can play no part in Carmeli's description of the large scale structure of the universe.

Even though it does not explicitly appear in Carmeli's \textit{spacevelocity} metric, the vacuum energy contribution to gravity $\Omega_{\Lambda}$ is a property of the metric and tends to unity as a function of decreasing redshift. This indicates that the universe, although always open because $\Omega_{m} < 1$, is asymptotically expanding towards a spatially flat state, with a value of $\Omega_{\Lambda} + \Omega_{m} \approx 1.021$ at the present epoch. 

\appendix*
\section{} 
This appendix deals with a more rigorous approach to the effect that curved \textit{spacevelocity} has on matter density as a function of redshift. However this approach results in a transcendental equation, which could not be used in analytical curve-fitting. Nevertheless the analysis here validates the methods used in sections \ref{sec:Comparison}, \ref{sec:density2}.

\begin{figure}
\includegraphics[width = 3.5 in]{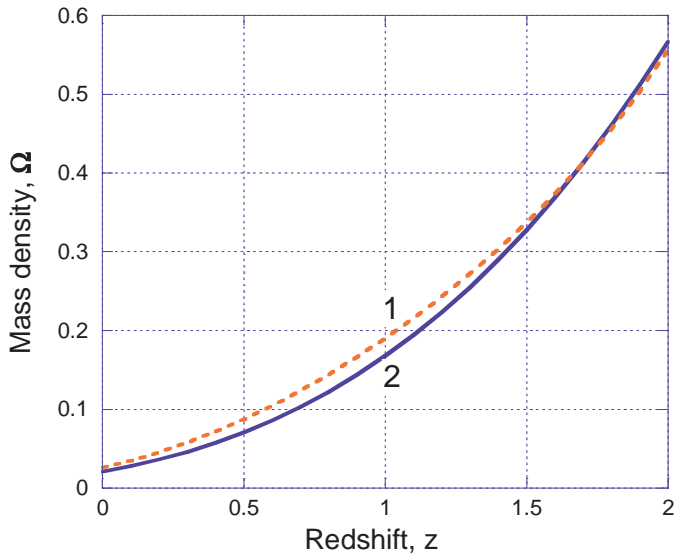}
\caption{\label{fig:figA1} (Color online) Mass density $\Omega$ as a function of redshift $z$ for the two models. Curve 2 (blue solid line) is $\Omega$ for the flat \textit{spacevelocity} model with $\Omega_{m} = 0.021$. Curve 1 (red dashed line) is $\Omega$ for the curved \textit{spacevelocity} model derived in the Appendix with $\Omega_{m} = 0.027$}
\end{figure}

If we start with the $z$-derivative of $\Omega$ from the right-hand side of (\ref{eqn:densityeqn}), we get
\begin{equation} \label{eqn:dOmega}
d\Omega = 3 \Omega_{m} (1+z)^{2} dz.
\end{equation} 
By the chain rule (\ref{eqn:dOmega}) can be written as
\begin{equation} \label{eqn:dOmegar}
d\Omega = 3 \Omega_{m} (1+z)^{2} \frac{dz}{dr} dr.
\end{equation}

In a flat \textit{spacevelocity} where $\Omega = 1$ and $e^{\xi/2} = 1$ it follows from (\ref{eqn:4Dmetricderivrz}) that $dz/dr = 1/c \tau$. Now using (\ref{eqn:4Dmetricderivrz}), in general,  
\begin{equation} \label{eqn:dOmegarcurved}
d\Omega = 3 \Omega_{m} (1+z)^{2} \frac{e^{\xi/2}}{c \tau} dr
\end{equation}
which when substituting $e^{\xi/2}$ from (\ref{eqn:soln1field2}) and (\ref{eqn:fvalue1}) becomes,
\begin{equation} \label{eqn:dOmegarcurved2}
d\Omega = 3 \Omega_{m}  \frac{(1+z)^{2}} {\sqrt{1+(1-\Omega)\left(\frac {r} {c \tau}\right)^{2}}} d\left(\frac{r}{c \tau} \right).
\end{equation}

Finally by substituting $r/c\tau = z$, which is the Hubble law to the lowest order in $z$ in the limit of no gravity, we arrive at an expression which is integrable and describes the matter density in the presence of curved \textit{spacevelocity}.
\begin{equation} \label{eqn:dOmegarcurved3}
d\Omega \approx 3 \Omega_{m}  \frac{(1+z)^{2}} {\sqrt{1+(1-\Omega)z^{2}}} dz.
\end{equation}

From (\ref{eqn:dOmegarcurved3}) it is clear we recover (\ref{eqn:dOmega}) when $\Omega = 1$. So all that needs to be done is integrate (\ref{eqn:dOmegarcurved3}) to get $\Omega$ as a function of $z$. But it is not simply solved. It appears difficult to separate $\Omega$ in the integrand on the right-hand side from $z$. So assuming that in  (\ref{eqn:dOmegarcurved3}) the value of $\Omega$ is approximately constant or only slowly changing as a function of $z$ we integrate and impose the boundary conditions that when $\Omega = 1$ and $z \ll 1$ then $\Omega = \Omega_{m} (1 + z)^{3}$. To satisfy both conditions in (\ref{eqn:dOmega}) the limits of integration must be from $-1$ to $z$. That is 
\begin{equation} \label{eqn:dOmegarcurved4}
\Omega \approx 3 \Omega_{m}  \int_{-1}^{z} \frac{(1+z')^{2}} {\sqrt{1+(1-\Omega)z'^{2}}} dz'.
\end{equation}
This results in
\begin{widetext}
\begin{equation} \label{eqn:dOmegarcurvedfinal}
\Omega \approx \frac{\Omega_{m}}{2(1-\Omega)^{3/2}} \left[ 3 \sqrt{1-\Omega} 
\left((4+z) \sqrt{1+(1-\Omega) z^{2}} - 3 \sqrt{2-\Omega} \right) + 
3(1 - 2 \Omega)  \left(arcsinh \sqrt{1 - \Omega} + arcsinh(z \sqrt{1 - \Omega})\right) \right],
\end{equation}
\end{widetext}
from which it may be determined that  $\Omega \rightarrow \Omega_{m} (1+z)^{3}$ for arbitary $z$ in the limit where $\Omega \rightarrow 1$. 

In the limit where $z \rightarrow 0$, $\Omega \rightarrow \Omega_{m} (1+z)^{3}$ where $\Omega \rightarrow 1$ and $\Omega \rightarrow \Omega_{m} (0.958 + 3z) \approx \Omega_{m} (1 + z)^{3}$ where $\Omega \rightarrow 0$. It is expected that the mass density can alway be approximated to yield Euclidean space locally. From the former, the boundary conditions are almost met and hence the approximation is reasonably valid.

The solution to (\ref{eqn:dOmegarcurvedfinal}) is not analytical and must be solved numerically. The form of the transendental equation does not lend itself to curve fitting as has been done in sections \ref{sec:Comparison}, \ref{sec:density2}. However, using the software package Mathematica, it is possible to create a function that is the solution to such an equation as (\ref{eqn:dOmegarcurvedfinal}). This was done and compared with the flat \textit{spacevelocity} model. The result is almost identical to (\ref{eqn:densityeqn}) for redshift $z \leq 2$, that is, when $\Omega_{m} = 0.021$ in the flat \textit{spacevelocity} model, $\Omega_{m} = 0.027$ in the resulting curved \textit{spacevelocity} model. See figure \ref{fig:figA1} where the density $\Omega$ has been plotted for both models up to $z = 2$.

This then means the matter density $\Omega(z)$ in the curved \textit{spacevelocity} model has approximately the same functional dependence on redshift $z$ used in the main text. Therefore we are justified in using the flat \textit{spacevelocity} model or the approximate curved \textit{spacevelocity} model. Only where $z > 2$ does the curve derived from (\ref{eqn:dOmegarcurvedfinal}) depart from the flat space model of (\ref{eqn:densityeqn}). Also for $z \gg 2$ the assumption, that $\Omega$ be at most a slowly changing function, breaks down and thus the integration in (\ref{eqn:dOmegarcurved4}) is invalid.

At these higher redshifts the unapproximated form of (\ref{eqn:dOmegarcurved2}) must be solved where $r/c \tau$ is substituted from (\ref{eqn:phasespacesolnnatural}) and $v/c \rightarrow \varsigma = ((1+z)^{2}-1)/((1+z)^{2}+1)$. But that is beyond the scope of this paper.

\end{document}